\documentclass[11pt]{article}
\usepackage{epsf,axodraw}
\def\beq{\begin{eqnarray}}
\def\eeq{\end{eqnarray}}
\def\bea{\begin{eqnarray*}}
\def\eea{\end{eqnarray*}}
\def\NPB#1#2#3{Nucl. Phys. {\bf B#1} (#2) #3}
\def\PLB#1#2#3{Phys. Lett. B {\bf #1} (#2) #3}
\def\PLBold#1#2#3{Phys. Lett. {\bf #1B} (#2) #3}
\def\PRD#1#2#3{Phys. Rev. D {\bf #1} (#2) #3}

\def\PRL#1#2#3{Phys. Rev. Lett. {\bf #1} (#2) #3}
\def\PREP#1#2#3{Phys. Rep. {\bf #1} (#2) #3}
\def\PTP#1#2#3{Prog. Theor. Phys. {\bf #1} (#2) #3}

\def\centeron#1#2{{\setbox0=\hbox{#1}\setbox1=\hbox{#2}\ifdim
\wd1>\wd0\kern.5\wd1\kern-.5\wd0\fi
\copy0\kern-.5\wd0\kern-.5\wd1\copy1\ifdim\wd0>\wd1
\kern.5\wd0\kern-.5\wd1\fi}}
\def\ltap{\>\centeron{\raise.35ex\hbox{$<$}}{\lower.65ex\hbox{$\sim$}}\>}
\def\gtap{\>\centeron{\raise.35ex\hbox{$>$}}{\lower.65ex\hbox{$\sim$}}\>}
\def\gsim{\mathrel{\gtap}}
\def\lsim{\mathrel{\ltap}}
\def\slashchar#1{\setbox0=\hbox{$#1$}           
   \dimen0=\wd0                                 
   \setbox1=\hbox{/} \dimen1=\wd1               
   \ifdim\dimen0>\dimen1                        
      \rlap{\hbox to \dimen0{\hfil/\hfil}}      
      #1                                        
   \else                                        
      \rlap{\hbox to \dimen1{\hfil$#1$\hfil}}   
      /                                         
   \fi}                                        %

\def\singleandabitspaced{\baselineskip=\normalbaselineskip\multiply
    \baselineskip by 117\divide\baselineskip by 100}
\def\singlespaced{\baselineskip=\normalbaselineskip}

\newcommand{\newc}{\newcommand}
\newc{ \GG         }{\tilde G}
\newc{ \Ni         }{ {\tilde N}_i }
\newc{ \Nj         }{ {\tilde N}_j }
\newc{ \NI         }{ {\tilde N}_1 }
\newc{ \NII        }{ {\tilde N}_2 }
\newc{ \NIII       }{ {\tilde N}_3 }
\newc{ \NIIII      }{ {\tilde N}_4 }
\newc{ \Ci         }{ {\tilde C}_i }
\newc{ \Cj         }{ {\tilde C}_j }
\newc{ \CI         }{ {\tilde C}_1 }
\newc{ \CII        }{ {\tilde C}_2 }
\newc{ \CIp        }{ {\tilde C}^{+}_1 }
\newc{ \CIpm       }{ {\tilde C}^{\pm}_1 }
\newc{ \CIm        }{ {\tilde C}^{-}_1 }
\newc{ \Cip        }{ {\tilde C}^{+}_i }
\newc{ \Cim        }{ {\tilde C}^{-}_i }
\newc{ \Cjp        }{ {\tilde C}^{+}_j }
\newc{ \Cjm        }{ {\tilde C}^{-}_j }
\newc{ \eL         }{ {\tilde e}_L }
\newc{ \eR         }{ {\tilde e}_R }
\newc{ \ser        }{ {\tilde e}_R }
\newc{ \smur       }{ {\tilde \mu}_R }
\newc{ \slr        }{ {\tilde \ell}_R }
\newc{ \slep        }{ {\tilde \ell} }
\newc{ \veL        }{ {\tilde \nu} }
\newc{ \dL         }{ \tilde d_L }
\newc{ \dR         }{ \tilde d_R }
\newc{ \uL         }{ \tilde u_L }
\newc{ \uR         }{ \tilde u_R }
\newc{ \slepton    }{ \widetilde \ell }
\newc{ \ltilde     }{ {\tilde \ell} }
\newc{ \nutilde    }{ {\tilde \nu} }
\newc{ \snu        }{ { \tilde \nu} }
\newc{ \stau       }{ { \tilde \tau} }
\newc{ \Et         }{ { \slashchar{E}_T } }
\newc{ \Etot       }{ { \slashchar{E} } }
\newc{ \spt        }{ { \slashchar{p}_T } }
\newc{ \GeV        }{ {\rm GeV} }

\begin{document}
\begin{titlepage}
\begin{flushright}
{\large
 hep-ph/9710217\\
 UM-TH-97-25\\
 SLAC-PUB-7668\\
 (revised version)
 }
 \end{flushright}
\vskip 1.2cm

\begin{center}
{\LARGE\bf Three-body decays of selectrons and smuons in}

{\LARGE\bf low-energy supersymmetry breaking models}

\vskip 2cm

{\large
 S.~Ambrosanio\footnote{Work supported mainly by 
an INFN postdoctoral fellowship, Italy. \\ 
\phantom{000} Address after Jan.~1, 1998:
Deutsches Elektronen-Synchrotron DESY, D-22603 Hamburg, Germany.},
 Graham D.~Kribs
} \\
\vskip 4pt
{\it Randall Physics Laboratory, University of Michigan,\\
     Ann Arbor, MI 48109--1120 } \\
\vskip 1.5mm 
{\large and}\\
\vskip 1.5mm 
{\large Stephen P.~Martin\footnote{Work supported 
by the Department of Energy under contract number DE-AC03-76SF00515. \\
\phantom{000} Address after Oct.~1, 1997:
Physics Department, University of California, Santa Cruz, CA 95064.}
} \\
{\it Stanford Linear Accelerator Center, Stanford University,\\
     Stanford, CA 94309 } \\

\vskip 2.0cm

\begin{abstract}
In models with low-energy supersymmetry breaking, it is well-known that
charged sleptons can be significantly lighter than the lightest
neutralino, with the gravitino and lighter stau being the lightest
and next-to-lightest supersymmetric particles respectively. We
give analytical formulas for the three-body decays of
right-handed selectrons and smuons into final states involving a tau, 
a stau, and an electron or muon, which are relevant in this scenario. 
We find that the three-body decays dominate 
over much of the parameter space, but the two-body decays into a lepton
and a gravitino can compete if the three-body phase space is small
and the supersymmetry-breaking scale (governing the two-body
channel) is fairly low. We study this situation quantitatively for
typical gauge-mediated supersymmetry breaking model parameters.
The three-body decay lengths are possibly macroscopic,
leading to new unusual signals. We also analyze the
final-state energy distributions, and briefly assess the 
prospects for detecting these decays at CERN LEP2 and other colliders.

\end{abstract}

\end{center}


\end{titlepage}
\setcounter{footnote}{0}
\setcounter{page}{2}
\setcounter{section}{0}
\setcounter{subsection}{0}
\setcounter{subsubsection}{0}

\newpage
\singleandabitspaced

\section*{1. Introduction}
\indent 

Supersymmetry-breaking effects in the Minimal Supersymmetric Standard
Model (MSSM) are usually introduced explicitly as soft terms
in the lagrangian. In a more complete theory, supersymmetry is
expected to be an exact local symmetry of the lagrangian which is
spontaneously broken in the vacuum state in a sector of particles distinct
from the MSSM. There are two main proposals for how supersymmetry
breaking is communicated to the MSSM particles. Historically, the more popular
approach has been that supersymmetry breaking occurs at a scale 
$\gsim 10^{10}$ GeV and is communicated to the
MSSM dominantly by gravitational interactions. In this case,
the lightest supersymmetric particle (LSP) is naturally the lightest
neutralino $(\NI )$. One of the virtues of this gravity-mediated
supersymmetry breaking scenario is that a neutralino LSP can easily
have the correct relic density to make up the cold dark matter with
a cosmologically acceptable density.

Recently, there has been a resurgence of interest in the 
idea \cite{oldGMSBmodels,newGMSBmodels} that supersymmetry-breaking
effects are communicated to the MSSM by the ordinary
$SU(3)_C \times SU(2)_L \times U(1)_Y$ gauge interactions rather than
gravity. This gauge-mediated supersymmetry breaking (GMSB) scenario
allows the ultimate supersymmetry-breaking order
parameter $\sqrt{F}$ to be much smaller than $10^{10}$ GeV, perhaps
even as small as $10^4 $ GeV or so, with the important
implication that the gravitino $(\GG)$ is the LSP. The spin-$3/2$
gravitino absorbs the would-be goldstino of supersymmetry breaking
as its longitudinal (helicity $\pm 1/2$) components by the super-Higgs
mechanism, obtaining a mass 
$ m_{\GG} = {F/ \sqrt{3} M_P} = 2.37 
( \sqrt{F}/ 100\> {\rm TeV} )^2 \> {\rm eV},
$
where $M_P = (8 \pi G_{\rm Newton} )^{-1/2} = 2.4 \times 10^{18}$ GeV
is the reduced Planck mass.
The gravitino inherits the non-gravitational interactions of the 
goldstino it has absorbed \cite{Fayet}. This means that
the next-to-lightest supersymmetric
particle (NLSP) can decay into its standard model partner and
a gravitino with a characteristic decay length which can be less than
of order 100 microns (for $\sqrt{F} \lsim 10^5$ GeV) or more than a 
kilometer
(for $\sqrt{F} \gsim 10^7$ GeV), or anything in between. This leads to
many intriguing phenomenological possibilities which are unique to
models of low-energy supersymmetry breaking
[3-10]. 
For kinematical
purposes, the gravitino is essentially massless.
The perhaps surprising
relevance of a light gravitino for collider physics
can be traced to the fact that
the interactions of the longitudinal components of the gravitino
are the same as those of the goldstino it has absorbed, and are proportional
to $1/m_{\tilde G}$ (or equivalently to $1/F$) in the light gravitino
(small $F$) limit \cite{Fayet}. 

In a large class of models with low-energy supersymmetry breaking,
the NLSP will either be the lightest neutralino or the lightest stau
$(\stau_1)$ mass eigenstate.
Our convention for the stau mixing angle $\theta_\stau$
is such that 
\beq
\pmatrix{\stau_1\cr \stau_2} = 
\pmatrix{\cos\theta_\stau & \sin\theta_\stau \cr 
         -\sin\theta_\stau & \cos\theta_\stau}
\pmatrix{\stau_L\cr \stau_R}
\label{staumix}
\eeq 
with 
$m_{\tilde \tau_1} < m_{\tilde \tau_2}$ 
and
$0\leq \theta_{\stau} < \pi$ (so $\sin\theta_\stau \geq 0$). 
The sign of $\cos\theta_\stau$ 
depends on the sign of $\mu$ (the superpotential Higgs mass parameter)
through the off-diagonal term $-\mu m_\tau \tan\beta$ in the
stau (mass)$^2$ matrix. This term typically dominates over the contribution
from the soft trilinear scalar couplings in GMSB models, because the
latter are very small at the messenger scale and because the effects of
renormalization group running are usually not very large.
For this reason, it is quite
unlikely that cancellation can lead to
$\cos\theta_\stau \approx 0$ in these models, unless
the scale of supersymmetry breaking is quite high.
In GMSB models like those in Ref.~\cite{AKMLEP2} which
are relevant to the decays studied in this paper,  
$|\cos\theta_\stau|$ ranges from
about $0.1$ to $0.3$ when the mass splittings between $\stau_1$ and the
lighter selectron and smuon are less than about 10 GeV. That is the
situation we will be interested in here. 
The selectrons and smuons also mix exactly analogously
to Eq.~(\ref{staumix}).  However, at least in GMSB models, their mixings
are generally much
smaller, with $\cos\theta_{\tilde \mu}/\cos\theta_{\tilde \tau}
\sim y_\mu/y_\tau \approx 0.06$ and $\cos\theta_{\tilde
e}/\cos\theta_{\tilde \tau}
\sim y_e/y_\tau \approx 3 \times 10^{-4}$. Therefore, in most cases
one can just treat the lighter selectron and smuon mass eigenstates
as nearly unmixed and degenerate states.
We will write these mass eigenstates as $\tilde e_R$ and $\tilde \mu_R$,
despite their small mixing. 
We will also assume, as is the
case in minimal GMSB models, that there are no lepton
flavor violating couplings or mixings.

The termination of superpartner decay chains depends crucially on the
differences between $m_{\NI}$, $m_{\stau_1}$, and $m_{\tilde \ell_R}$
(in this paper $\ell $ is generic notation for $e$ or $\mu$). We assume
that \mbox{$R$-parity} violation
is absent, so that there are no competing decays for the NLSP.
If the NLSP is $\NI$ with $m_{\NI} < m_{\stau_1} - m_\tau$, then the
decay $\NI \rightarrow \gamma\GG$ can lead to new discovery signals for
supersymmetry, as explored in 
Refs.~[3-9].
In other models, one finds that the NLSP is $\stau_1$
\cite{DDRT}. Here one 
must distinguish
between several qualitatively distinct scenarios. If $\tan\beta$
is not too large, then $\ser$ and $\smur$
will not be much heavier than $\stau_1$, and the decays 
$\slr \rightarrow \ell \tau \stau_1$ and $\slr \rightarrow \ell \NI$
will not be kinematically open.
In this ``slepton co-NLSP scenario'',
each of $\ser$, $\smur$, and $\stau_1$ may decay according to
$\ser \rightarrow e \GG$, $\smur \rightarrow \mu \GG$ and $\stau_1
\rightarrow
\tau \GG$, possibly with very long lifetimes. 
There can also be competing three-body decays $\slr \rightarrow \nu_\ell
\overline \nu_\tau \stau_1$ through off-shell charginos ($\tilde C_i$).
However, these decays are strongly suppressed by phase space
and because the coupling of $\slr$ to 
$\nu_\ell \tilde C_i$ is very small. In the approximations that $
m_{\slr}^2 - m_{\stau_1}^2 \ll m_{\CI}^2$ 
and $1 - m^2_{\stau_1}/m^2_{\slr} \ll 1$, one finds
\beq
\Gamma
(\slr^- \rightarrow \nu_\ell \overline \nu_\tau \stau^-_1) =
{\alpha^2 m_{\slr}\over 960 \pi \sin^4\theta_W} 
\left (1-{m_{\stau_1}^2\over m_{\slr}^2} \right )^5
\sum_{i,j=1,2}
{d_i^{\tilde \ell *} d_j^{\tilde \ell }
d_i^{\tilde \tau} d_j^{\tilde \tau *}
\over 
(m^2_{\tilde C_i}/m_{\slr}^2 - 1)
(m^2_{\tilde C_j}/m_{\slr}^2 - 1)
}
\label{nunutauwidth} 
\eeq
where 
$d_i^{\tilde f} = U_{i1} \cos \theta_{\tilde f} - (y_f/g)
U_{i2} \sin\theta_{\tilde f}$, with Yukawa couplings  
$y_f = g m_f/(\sqrt{2} m_W
\cos\beta )$, for $f=\ell,\tau$. 
Here $U_{ij}$ is one of the chargino mixing matrices in the notation 
of \cite{conventions} and $g$ is the $SU(2)_L$ gauge coupling. 
(Of course, the decay $\slr^+ \rightarrow  \overline\nu_\ell
\nu_\tau \stau^+_1$ has the same width.) For $\smur$ decays, we find that
this width is always less than about $10^{-7}$ eV 
in GMSB models like the ones discussed in
\cite{AKMLEP2} if
$m_{\smur} - m_{\stau_1} < m_\tau$ and $m_{\tilde \mu_R} \gsim 80$ GeV.
The maximum width decreases with increasing $m_{\tilde \mu_R}$
as long as we continue to require that the decay $\tilde \mu_R 
\rightarrow \mu \tau\stau_1$ is not kinematically open. 
(For the corresponding $\ser $
decays, the width is more than four orders of magnitude smaller.) 
This corresponds to 
physical decay lengths of (at least) a few meters unless the
sleptons are produced very close to threshold. It is possible to
have somewhat enhanced widths if $m_{\CI} - m_{\stau_1}$ is 
decreased or if $\cos\theta_{\tilde \mu}$ is increased compared to
the values typically found in GMSB models. However, even if
the decays
$\slr \rightarrow \nu_\ell \overline\nu_\tau \stau_1$ can occur within
a detector, they will be extraordinarily hard to detect because
the neutrinos are unobserved and the $\stau_1$ momentum in the lab
frame will not be
very different from that of the decaying $\slr$. 
The subsequent decays $\stau_1 \rightarrow \tau\GG$ can be
distinguished from the direct $\slr \rightarrow \ell\GG$, but if the
latter can occur within the detector, then they will likely
dominate over
$\slr \rightarrow \nu_\ell \overline \nu_\tau \stau_1$ anyway. So
it is very  doubtful
that the decays $\slr \rightarrow \nu_\ell \overline\nu_\tau \stau_1$ can
play a role in collider phenomenology.

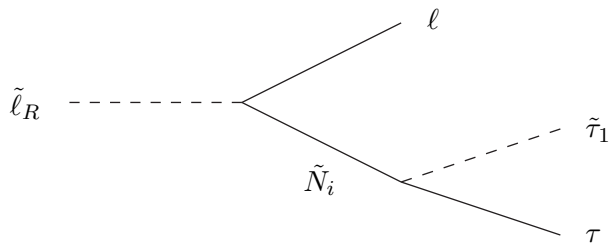
\begin{figure}[tb]
\begin{picture}(400,100)(0,0)
   \DashLine(  105, 60 )( 170, 60 ){4}
   \Line( 170, 60 )( 230, 90 )
   \Line(      170, 60 )( 230, 30 )
   \Line( 290, 10 )( 230, 30 )
   \DashLine(  230, 30 )( 290, 50 ){4}
   \Text(       98, 60 )[r]{ $\tilde{\ell}_R$ }
   \Text(      200, 31 )[c]{ $\tilde{N}_i$ }
   \Text(      237, 92 )[l]{ $\ell$ }
   \Text(      297, 50 )[l]{ $\tilde{\tau}_1$ }
   \Text(      297, 10 )[l]{ ${\tau}$ }
\end{picture}
\caption{Right-handed selectrons and smuons can decay according to
$\slr^- \rightarrow \ell^- \tau^+ \stau_1^-$ or
$\slr^- \rightarrow \ell^- \tau^- \stau_1^+$, with different
matrix elements,
through virtual neutralinos
$\Ni$ ($i=1,2,3,4$). 
}
\label{figthreebodyfeynman}
\end{figure}

For larger values of
$\tan\beta$, enhanced stau mixing renders 
$\tilde \tau_1$ lighter than $\ser$ and $\smur$
by more than $m_\tau$. In this ``stau NLSP scenario",
all supersymmetric decay chains should (naively)
terminate in $\stau_1 \rightarrow \tau \GG$ \cite{DDRT, DDN, AKMLEP2}, 
again possibly with a 
very long lifetime.\footnote{An important exception occurs if 
$|m_{\stau_1} - m_{\NI}| < m_\tau $ and $m_{\NI} < m_{\slr}$.
In this ``neutralino-stau co-NLSP scenario", both of the decays
$\stau_1 \rightarrow \tau \GG$ and $\NI \rightarrow \gamma \GG$
occur without significant competition.} If the mass ordering is
$m_{\smur} - m_\mu$ and/or 
$m_{\ser} - m_e > m_{\NI}$, 
then the two-body decays
$\smur \rightarrow \mu \NI$ and/or $\ser \rightarrow e \NI$ will be open
and will dominate. 
In the rest of this paper, we will instead consider the situation in the
stau NLSP scenario in which 
$m_{\NI} > m_{\slr} - m_\ell > m_{\stau_1} + m_\tau$. In that case,
$\smur$ and/or $\ser$ can decay through off-shell neutralinos in
three-body modes $\smur \rightarrow \mu \tau \stau_1$ and/or 
$\ser \rightarrow e \tau \stau_1$, as shown in 
Fig.~\ref{figthreebodyfeynman}.
Here one must be careful to distinguish between the 
different charge channels $\tau^+\stau_1^-$ and 
$\tau^-\stau_1^+$ in the final state, for a given
charge of the decaying slepton. In the following we will give formulas
for $\Gamma(\slr^- \rightarrow \ell^- \tau^+ \stau_1^-)$ and
$\Gamma(\slr^- \rightarrow \ell^- \tau^- \stau_1^+)$, which in 
general can be quite different,\footnote{We are indebted to Nima
Arkani-Hamed for pointing this out to us.} 
except when the virtual neutralino is nearly on shell.
[Of course, these
are equal to 
$\Gamma(\slr^+ \rightarrow \ell^+ \tau^- \stau_1^+)$ and
$\Gamma(\slr^+ \rightarrow \ell^+ \tau^+ \stau_1^-)$, respectively.]
These three-body slepton decays have 
been rightly ignored in 
previous phenomenological discussions of the MSSM with a neutralino LSP,
in which the two-body decays $\slr \rightarrow \ell \NI$ (and possibly
others) are always open. However, in models with a gravitino LSP,
$\NI$ is allowed to be much heavier, so it is important to realize
that three-body decays of $\ser$ and $\smur$ are relevant and
can in principle imply long lifetimes and macroscopic decay lengths. 
In the following, we will present analytical results for the three-body
decay widths of $\ser$ and $\smur$, and study numerical results for
typical relevant model parameters.

\section*{2. Three-body slepton decay widths}
\indent 

Let us first consider the ``slepton-charge preserving" decay
$\slr^- \rightarrow \ell^- \tau^+ \stau_1^-$, keeping
$\slep$ mixing effects.
The matrix element for the 
relevant
Feynman diagrams in Fig.~1 can be written as
\beq
{\cal M}
= \sum_{j=1}^4 {\overline u}(p_{\ell}) 
( P_R a_j^{\tilde \ell *} + P_L b_j^{\tilde \ell *} )
{ 
\slashchar{p}_{\ell} - \slashchar{p}_{\slr}+ m_{\Nj}  
\over
  [ m_{\Nj}^2 - m_{\slr}^2 (1-x_\ell+r_\ell^2) ] } 
( P_L a_j^{\tilde \tau } + P_R b_j^{\tilde \tau } )
v(p_\tau) ,
\label{matrixele}
\eeq
where $P_{L,R} = (1\pm \gamma_5)/2$, and 
$x_\ell = 2 p_{\slr} \cdot p_\ell/p^2_{\slr}$, and
\beq
a^{\tilde \tau}_j &=& \sqrt{2} g^\prime
N_{j1} \sin\theta_{\stau}
+ y_\tau N_{j3} \cos\theta_{\stau}; 
\\ b_j^{\tilde \tau} &=& -{1\over \sqrt{2}}
(g^\prime N^*_{j1} + g N_{j2}^*) \cos\theta_{\stau}
+ y_\tau N^*_{j3} \sin\theta_{\stau},
\eeq
with exactly analogous formulas for $a_j^{\tilde \ell}$
and $b_j^{\tilde \ell}$, with $\tilde \tau \rightarrow \tilde \ell$.
Here we have 
adopted the notation of Ref.~\cite{conventions} for the 
unitary (complex) neutralino mixing matrix $N_{ij}$ with all $m_{\tilde N_i}$
real and positive, and 
$g$ and $g^\prime$ are the $SU(2)_L$ and $U(1)_Y$ gauge couplings.
Our fermion propagator is
proportional to $(-\slashchar{p} + m)/(p^2 + m^2)$, with a spacetime
metric signature $(-\mbox{+++})$. 

Summing over final state
spins 
and performing the phase space
integration, we obtain:\footnote{Similar formulas can be derived for 
the three-body decay widths of all sfermions in the MSSM. Here we have
neglected
higher order effects, including contributions to the neutralino widths
from final states other than $\tau\stau_1$, since we will be interested
in the situation in which $m_{\NI}$ is not too close to $m_{\slr}$.}
\beq
\Gamma(\slr^- \rightarrow \ell^- \tau^+ \stau^-_1) = 
{m_{\slr} \over 512 \pi^3 } \sum_{i,j=1}^4 
\sum_{a=1}^6 c^{(a)}_{ij} I^{(a)}_{ij}
\label{fullwidth}
\eeq
in terms of coefficients
\beq
&&c^{(1)}_{ij} = a_j^{\slep *} a_j^\stau a_i^{\slep} a_i^{\stau *} +
            b_j^{\slep *} b_j^\stau b_i^{\slep} b_i^{\stau *},
\\
&&c^{(2)}_{ij} = 
a_j^{\slep *} b_j^\stau a_i^{\slep} b_i^{\stau *}
+ b_j^{\slep *} a_j^\stau b_i^{\slep} a_i^{\stau *} ,
\\
&&c^{(3)}_{ij} = 2{\rm Re}[ 
a_j^{\slep *} b_j^\stau a_i^{\slep} a_i^{\stau *} 
+ b_j^{\slep *} a_j^\stau b_i^{\slep} b_i^{\stau *} ] ,
\\
&&c^{(4)}_{ij} = -2{\rm Re}[ 
            a_j^{\slep *} b_j^\stau b_i^{\slep} b_i^{\stau *}  +
b_j^{\slep *} a_j^\stau a_i^{\slep} a_i^{\stau *} ],
\\
&&c^{(5)}_{ij} = -4{\rm Re}[ b_j^{\slep *} a_j^\stau a_i^{\slep} b_i^{\stau 
*} ], \\
&&c^{(6)}_{ij} = -4{\rm Re}[ a_j^{\slep *} a_j^\stau b_i^{\slep} b_i^{\stau 
*} ], \eeq
and
dimensionless integrals $I_{ij}^{(a)}$ defined
as follows. First, we introduce the mass ratios
$r_{\stau} = m_{\stau_1}/m_{\slr}$, 
$r_{\tau} = m_{\tau}/m_{\slr}$, 
$r_{\ell} = m_{\ell}/m_{\slr}$, and  
$r_{\Ni} = m_{\Ni}/m_{\slr}$ 
with
$r_\ell \ll r_\tau \ll r_{\stau} < 1 - r_\tau - r_\ell  
$ and $ r_{\NI} > 1 - r_\ell$.
Then we find
\beq
I_{ij}^{(1)} &=& \int dx_\ell \>\, 
(x_\ell - 2 r_\ell^2)
(1-x_\ell+r_\ell^2 ) 
(1-x_\ell + r_\ell^2 + r_\tau^2 -
r_{\stau}^2)
f_{ij},
\\
I_{ij}^{(2)} &=& r_{\Ni} r_{\Nj}
\int dx_\ell \>\, 
(x_\ell - 2 r_\ell^2)(1-x_\ell + r_\ell^2 + r_\tau^2 -
r_{\stau}^2)f_{ij},
\\
I_{ij}^{(3)} &=& 2 r_{\tau} r_{\Nj}
\int dx_\ell\>\,
(x_\ell - 2 r_\ell^2) 
(1-x_\ell + r_\ell^2) f_{ij},
\\
I_{ij}^{(4)} &=& 2 r_{\ell} r_{\Nj}
\int dx_\ell \>\, 
(1-x_\ell + r_\ell^2)(1-x_\ell + r_\ell^2 + r_\tau^2 -
r_{\stau}^2)f_{ij},
\\
I_{ij}^{(5)} &=& 2 r_{\ell} r_{\tau} r_{\tilde N_i} r_{\tilde N_j}
\int dx_\ell\>\,
(1-x_\ell + r_\ell^2) f_{ij},
\\
I_{ij}^{(6)} &=& 2 r_{\ell} r_{\tau}
\int dx_\ell\>\,
(1-x_\ell + r_\ell^2)^2 f_{ij},
\eeq
where
\beq
f_{ij} = {
\sqrt{x_\ell^2 - 4 r_\ell^2} \>\lambda^{1/2}[1-x_\ell+r_\ell^2,
r_\tau^2, r_{\stau}^2] \over (1-x_\ell+r_\ell^2)^2\, 
(r_{\Ni}^2 -1 + x_\ell - r_\ell^2)
\,(r_{\Nj}^2 -1 + x_\ell - r_\ell^2)
}
\label{deffij}
\eeq 
with $\lambda^{1/2}[a,b,c] = \sqrt{a^2 + b^2 + c^2 - 2 a b - 2 a c - 2 bc}$.
The limits of integration for $x_\ell$ are $2 r_\ell < x_\ell <
1+r_\ell^2 - r_\tau^2 - r_{\stau}^2 - 2 r_\tau r_{\stau}$.
The matrix element and decay width for the ``slepton-charge flipping" channel
$\slr^- \rightarrow \ell^- \tau^- \stau_1^+$ are obtained by replacing 
$a_j^{\tilde \tau} \rightarrow b_j^{\tilde \tau *}$ and
$b_j^{\tilde \tau} \rightarrow a_j^{\tilde \tau *}$ everywhere in 
the above equations. 

In GMSB models like those studied in 
Ref.~\cite{AKMLEP2} which are relevant to these decays, one
finds that $m_{\ser} - m_{\smur}$ is at the most a few tens of MeV,
so we will neglect the distinction between $m_{\ser}$ and $m_{\smur}$.
It is an excellent approximation to take 
$r_\mu=0$ {\it except} when
the mass difference 
\beq
\Delta m = m_{\slr} - m_{\stau_1} - m_{\tau}
\eeq 
is
a few hundred MeV or less, and $r_e=0$ is of course nearly always
a good approximation. 
It is also generally an excellent
approximation to neglect smuon and selectron mixing and Yukawa couplings
in the matrix element,
so that $a_j^{\tilde \ell}\approx \sqrt{2} g^\prime N_{j1}$ and
$b_j^{\tilde \ell} \approx 0$.\footnote{
We have calculated 
the effect of including the smuon mixing and the muon
Yukawa to be at the
level of a few to ten percent of the total smuon width,
for typical GMSB models from Ref.~\cite{AKMLEP2}.}
The effects of $I_{ij}^{(4,5,6)}$ are usually quite negligible because of
the $r_{\ell}$ and $b_j^{\slep}$ suppressions.
An instructive limit which is often approximately realized in GMSB
models (or, in generic models with gaugino mass unification, whenever 
$|\mu|$ is sizeably larger than the gaugino mass parameters) is the case in 
which the contributions from a Bino-like
$\NI$ dominate, with 
$m_{\NI} \approx 0.5 m_{\NII} \ll m_{\NIII}, m_{\NIIII}$. Since the
decaying $\slr$ essentially couples only to the Bino ($\tilde{B}$) 
component of the virtual neutralinos, this approximation is quite
good for a large class of models where $|N_{11}|$ is not too far from 1.
In that case, we may
neglect the contributions of virtual $\tilde N_2$, $\tilde N_3$ and 
$\tilde N_4$ 
because of the coupling constant
suppressions together with the suppressions due to larger neutralino
masses. With these approximations,
the expressions for the decay widths simplify to
\beq
&&
\Gamma(\slr^- \rightarrow \ell^-\tau^+\stau^-_1) \approx 
{m_{\slr} \over 512 \pi^3} \left [
|A_1|^2  I_{11}^{(1)} 
+ |B_1|^2 I_{11}^{(2)} 
- 2 {\rm Re}(A_1 B_1^*)\, I_{11}^{(3)} 
\right ],
\label{approxwidth}
\\
&&
A_1 = 2 g^{\prime 2} |N_{11}|^2 \sin\theta_\stau
+ \sqrt{2} g^\prime y_\tau N_{11}^* N_{13} \cos\theta_{\stau},
\\
&&
B_1 = g^{\prime 2} N_{11}^{*2} \cos\theta_\stau 
+ g g^\prime N^*_{11} N_{12}^* \cos\theta_{\stau} -
\sqrt{2} g^\prime y_\tau N^*_{11} N^*_{13} \sin\theta_\stau ,
\label{approxwidthAB}
\eeq
and
\beq
&&\Gamma(\slr^- \rightarrow \ell^-\tau^-\stau^+_1) \approx 
{m_{\slr} \over 512 \pi^3} \left [
|B^\prime_1|^2  I_{11}^{(1)} 
+ |A^\prime_1|^2 I_{11}^{(2)} 
- 2 {\rm Re}(A^\prime_1 B_1^{\prime *})\, I_{11}^{(3)} 
\right ],
\label{approxwidthother}
\\
&&
A^\prime_1 = 2 g^{\prime 2} N_{11}^2 \sin\theta_\stau
+ \sqrt{2} g^\prime y_\tau N_{11} N_{13} \cos\theta_{\stau},
\\
&&
B^\prime_1 = g^{\prime 2} |N_{11}|^{2} \cos\theta_\stau 
 + gg^\prime N_{11} N_{12}^* \cos\theta_{\stau} -
\sqrt{2} g^\prime y_\tau N_{11} N^*_{13} \sin\theta_\stau .
\label{approxwidthABother}
\eeq

We will be interested in the situation in which
$\Delta m$ is small (less than 10 GeV).
This implies 
that $\tan\beta$ is not too large,\footnote{For example in the GMSB models
studied in \cite{AKMLEP2} with $0 < \Delta m < 10$ GeV, the relevant range 
for $\tan\beta$ is from 
about 5 to 20 for sleptons that could be accessible at LEP2 or Tevatron
upgrades, with smaller values of $\tan\beta$ corresponding to smaller 
$\Delta m$.}
and 
thus $\stau_1$ has a large $\stau_R$ content. However, as we will
see in the next section, it
is usually {\it not} a good approximation to neglect stau mixing altogether
(by setting $\sin\theta_{\stau}=1$, $\cos\theta_{\stau} = 0$
everywhere), because $|
\cos\theta_{\stau} |$ is
likely to be at least $0.1$ as we have already mentioned. 
Near threshold, the range of integration includes only small values
of $x_\ell$, so that the dimensionless integrals 
$I_{11}^{(1)}$ and
 $I_{11}^{(2)}$ and $I_{11}^{(3)}$ scale approximately proportional to
$1/(r_{\NI}^2-1)^2$
and $r^2_{\NI}/(r_{\NI}^2-1)^2$ and
$r_{\NI}/(r_{\NI}^2-1)^2$, respectively. This means that the decay
width is suppressed as $r_{\NI}$ (or equivalently $m_{\NI}$)
is increased, with other parameters
held fixed. This is particularly likely in GMSB models with a 
large messenger sector and a high scale of supersymmetry 
breaking. Furthermore, the relative sizes of the $I^{(2)}_{11}$ and
$I^{(3)}_{11}$ contributions are enhanced in the large $r_{\NI}$
limit. 
It is important to note that as $r_{\NI}$ is increased, 
$\Gamma(\slr^- \rightarrow \ell^- \tau^- 
\stau_1^+)$ becomes larger than  $\Gamma(\slr^- \rightarrow \ell^- 
\tau^+ \stau_1^-)$, because of this effect together with the
fact that $A_1$ and 
$A_1^\prime$ typically have larger magnitudes than $B_1$ and $B_1^\prime$.
 Note also that the $I^{(3)}_{11}$ contributions appear to be 
suppressed
by a factor of $r_{\tau}$, but this turns out to be illusory since 
near threshold
$m_{\tau}$ is not the only small mass scale in the problem; in
particular it can be comparable to or even much larger than 
$\Delta m - m_{\ell}$ which
determines the kinematic suppression of the decay.

\section*{3. Numerical results}
\indent 

\begin{figure}[tb]
\centering
\epsfxsize=3.8in
\hspace*{0in}
\epsffile{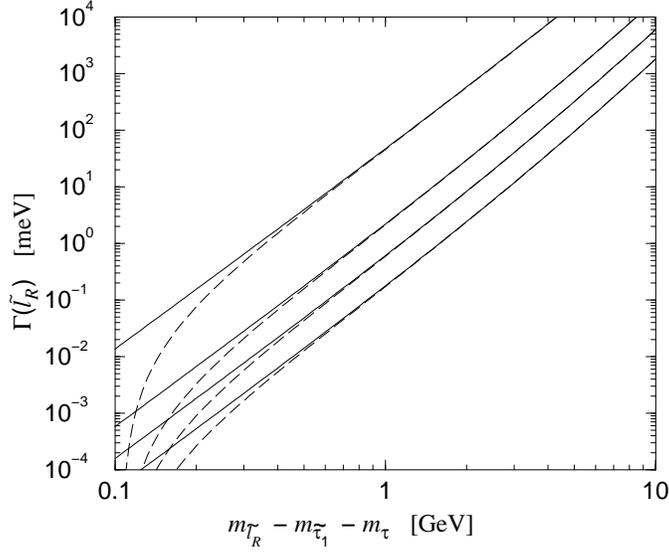}
\caption{
The decay widths in meV for $\ser \rightarrow e\tau\stau_1$ (solid lines)
and $\smur \rightarrow \mu\tau\stau_1$ (dashed lines), including
both $\tau^+\stau^-_1$ and $\tau^-\stau^+_1$ final states,
as functions of $\Delta m = m_{\slr} - m_{\stau_1} - m_{\tau}$.
The widths have been computed using
Eqs.~(\ref{approxwidth})-(\ref{approxwidthABother})
with $m_{\slr} = 90$ GeV, 
$\cos\theta_\stau = 0.15$ 
and
$r_{\NI} = 1.1$, 1.5, 2.0, and 3.0 (from top to bottom), with the
approximation $N_{11} = 1$ and $N_{12} = N_{13} = 0$.
}
\label{figthreebody90}
\end{figure}

\begin{figure}[tb]
\centering
\epsfxsize=3.3in
\hspace*{0in}
\epsffile{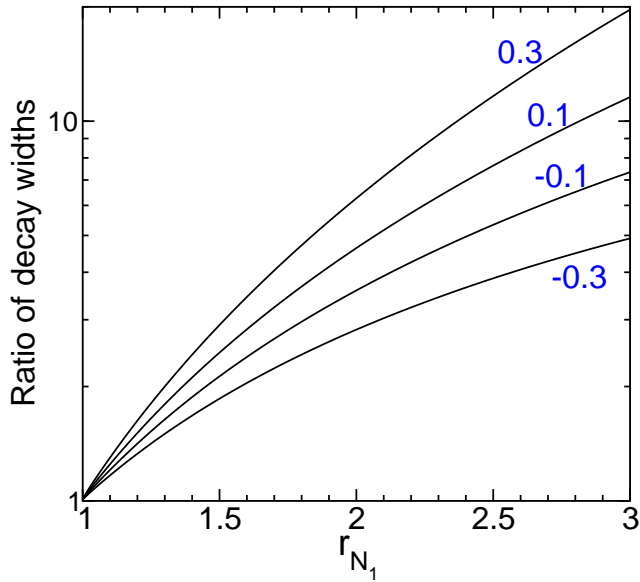}
\caption{
The ratio  $\Gamma(\tilde e_R^- \rightarrow e^- \tau^- \stau_1^+)/
\Gamma(\tilde e_R^- \rightarrow e^- \tau^+ \stau_1^-)$ is shown as 
a function of $r_{\NI}$, for four values of $\cos\theta_{\stau} =$
$0.3$, $0.1$, $-0.1$ and $-0.3$ (from top to bottom).
The widths have been computed using
Eqs.~(\ref{approxwidth})-(\ref{approxwidthABother})
with $m_{\tilde e_R} = 90$ GeV, and with the
approximation $N_{11} = 1$ and $N_{12} = N_{13} = 0$. 
}
\label{figratio}
\end{figure}

\begin{figure}[tb]
\centering
\epsfxsize=3.3in
\hspace*{0in}
\epsffile{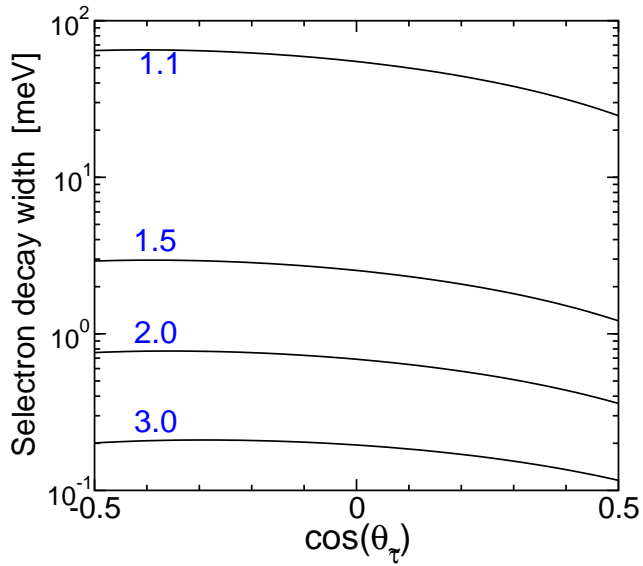}
\caption{
The dependence of $\Gamma(\ser \rightarrow e \tau\stau_1)$
(including both charge final states)
on $\cos\theta_{\stau}$, computed as in Fig.~\ref{figthreebody90} but 
with $\Delta m = m_{\ser} - m_{\stau} - m_\tau$ held fixed at 1.0 GeV. 
The four curves correspond to
$r_{\NI} = 1.1$, 1.5, 2.0, and 3.0 (from top to bottom).
}
\label{figvaryangle}
\end{figure}

\begin{figure}[tb]
\centering
\epsfxsize=3.8in
\hspace*{0in}
\epsffile{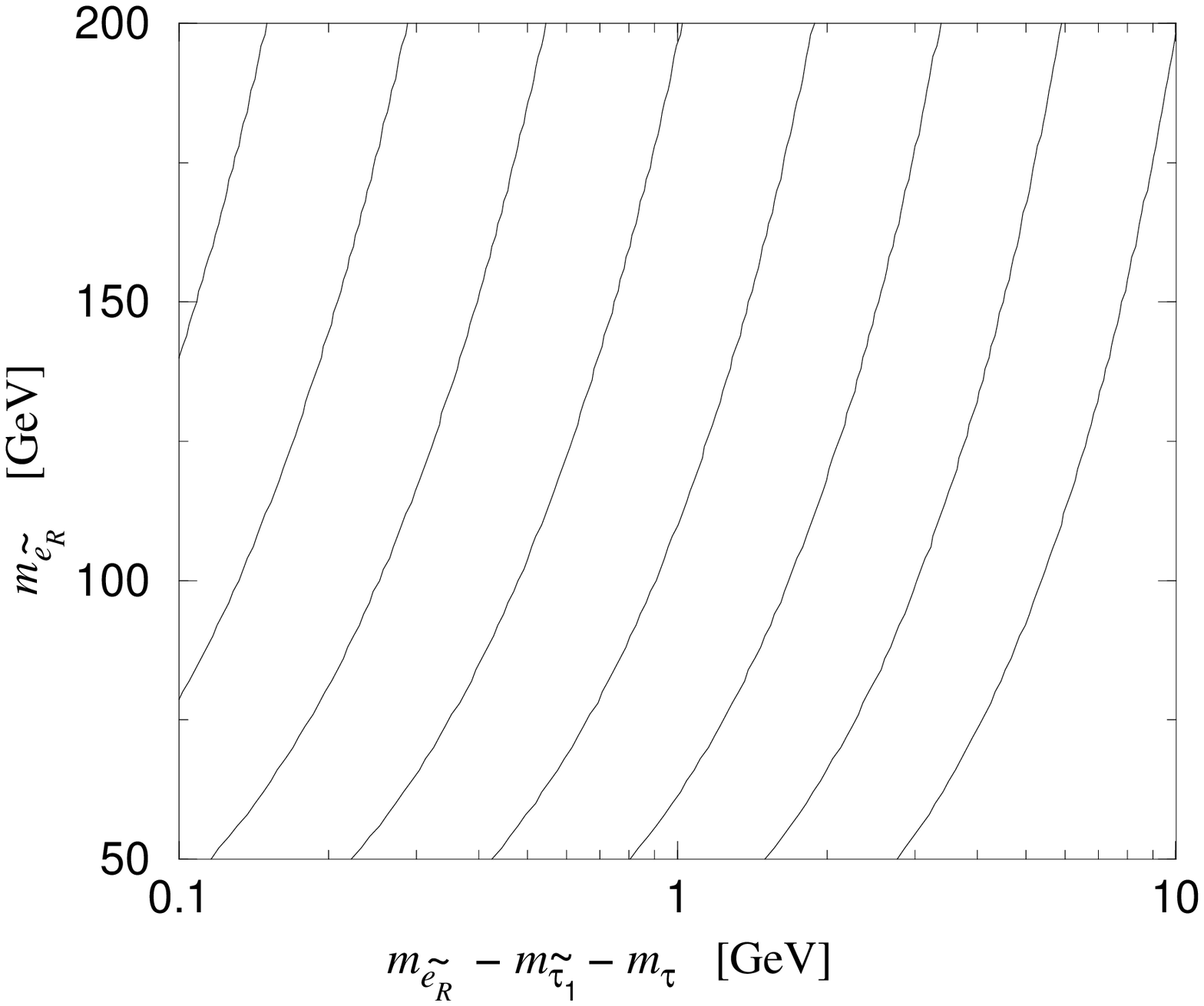}
\caption{
Contours of constant total decay width $\Gamma(\ser \rightarrow 
e\tau\stau_1)$ (from left to right, 0.0001, 0.001, 0.01, 0.1, 1,
10, 100 and 1000 meV), 
including both charge channels for the final 
state, 
and computed with $r_{\NI} = 1.5$ and 
$\cos\theta_\stau = 0.15$ with the same approximations as in 
Figs.~\ref{figthreebody90}-\ref{figvaryangle}.
}
\label{figcontours}
\end{figure}

Some typical results are shown in 
Figs.~\ref{figthreebody90}-\ref{figcontours}. In 
Fig.~\ref{figthreebody90},
we give the total three-body decay widths
for $\ser$ and $\smur$ (including both $\tau^+\stau^-_1$ and 
$\tau^-\stau^+_1$ final states)
as functions of $\Delta m$
for $m_{\slr} = 90$ GeV and four choices $r_{\NI} = 1.1$, 1.5, 2.0, and 3.0.
(In the GMSB models studied in Ref.~\cite{AKMLEP2}, one finds 
$r_{\NI} \lsim 
1.8$, but it is possible to imagine more general models with larger values.)
Here we have chosen the approximation of
Eqs.~(\ref{approxwidth})-(\ref{approxwidthABother}) with
$N_{11} = 1$ and
$\cos\theta_{\stau} = 0.15$. 
We use $m_{\tau} = 1.777$ GeV, $m_{\mu} = 0.1057$ GeV, 
$\alpha = 1/128.0$, and $\sin^2\theta_W = 0.2315$. Realistic
model parameters can introduce a significant variation in the decay
widths, and in general one should use the full formulas
given above
for any specific model. Our
choice of
a positive value for $\cos\theta_\stau$ in this example leads to a
suppression
in the width compared to the opposite choice, because of the sign of
the interference terms proportional to $I_{11}^{(3)}$ in 
Eqs.~(\ref{approxwidth}) and (\ref{approxwidthother}). 
These interference
terms are often of the order of tens of percent of the total width,
showing the importance of keeping the stau mixing effects if real
accuracy is needed.

The important ratio of the partial widths for the two charge 
channels $\Gamma(\slr^- \rightarrow \ell^- \tau^- 
\stau_1^+)/\Gamma(\slr^- \rightarrow \ell^- \tau^+ \stau_1^-)$ is 
shown in Fig.~\ref{figratio} for the case $\ell=e$, as a function of 
$r_{\NI}$. Here we have chosen  
values of $\cos\theta_{\stau} = -0.3$, $-0.1$, $0.1$ and $0.3$, and other
parameters as in Fig.~\ref{figthreebody90}. As expected,
this ratio is close to 1 when the virtual neutralino is nearly
on-shell, and increases with $r_{\NI}$. It scales roughly like
$I^{(2)}_{11}/I^{(1)}_{11} \approx r_{\NI}^2$, up to significant
corrections from the interference term(s). This increase  
tends to be 
more pronounced for larger $\cos\theta_{\stau}$
in these models. 
Because large $r_{\NI}$ also corresponds to longer lifetimes,
the decay $\slr^- \rightarrow \ell^- \tau^- \stau_1^+$ is likely to
dominate if the three-body decay lengths are macroscopic.

The variation with the stau mixing angle is 
further illustrated in 
Fig.~\ref{figvaryangle}, 
where we show the total three-body decay width
$\Gamma(\ser \rightarrow e\tau\stau_1)$ including
both charge final states with $m_{\ser} = 90$ GeV
and $\Delta m  = 1.0$ GeV, $r_{\NI} = 1.1$, 1.5, 2.0, 3.0, for the range 
$-0.5 < \cos\theta_{\stau} < 0.5$.
Note that 
the total width can vary by 
a factor of two
or more over this range.
Here it should be kept in mind that at least in the GMSB models studied
in Ref.~\cite{AKMLEP2}, one finds $0.1 \lsim |\cos\theta_\stau| \lsim
0.3$, so that the whole range shown may not be relevant. 
In Fig.~\ref{figcontours}, we show contours of constant 
total three-body decay widths
$\Gamma(\ser \rightarrow e \tau\stau_1)$
in the $\Delta m$ vs.~$m_{\ser}$ plane, for the choice $r_{\NI} = 1.5$ and 
$\cos\theta_\stau = 0.15$.
In both figures we continue to use $N_{11} = 1$, 
$N_{12} = N_{13} = 0$ in
Eqs.~(\ref{approxwidth})-(\ref{approxwidthABother}).
However, it should be emphasized
that in realistic models 
the
effects of deviations from this simplistic approximation can be quite
appreciable, especially since 
$|N_{11}|^2$ can easily be of order 0.7 or somewhat less in GMSB models,
and 
the width scales essentially like
$|N_{11}|^4$.

As can be seen from these figures, the physical 
three-body decay lengths
for $\ser$ and
$\smur$ can be quite large if
$\Delta m$ is less than
a few GeV and/or $m_{\NI}/m_{\slr}$ is large. In the lab frame,
the probability that a slepton $\slr$ with energy $E$ will travel
a distance $x$ before decaying is $P(x) = e^{-x/L}$, where
\beq
L = 0.2 \left ({\Gamma \over 1\>{\rm meV}}
\right )^{-1} \Bigl ( {E^2\over m^2_{\slr}}-1\Bigr )^{1/2}
{\rm mm}.
\label{defL}
\eeq
For sleptons pair-produced at LEP2 (or at a next-generation
lepton 
collider), $E$ in Eq.~(\ref{defL}) is simply
the beam energy. So if $\Delta m$
is less than a GeV or so (depending on $r_{\NI}$ and the specific
couplings of the model), $\ser$ and $\smur$ could have a macroscopic
and measurable decay length. If $\Delta m$ is of order 100
MeV or less, the decay length could even exceed the dimensions of 
typical detectors.

It is also important to realize that the dominant
decay for $\slr$ is not {\it a priori} known, since 
the three-body decays $\slr \rightarrow
\ell\tau\stau_1$ 
have to compete with the two-body decays to the gravitino $\slr \rightarrow
\ell\GG$. 
The latter have a width given by
\beq
\Gamma(\slr \rightarrow \ell\GG) = {m_{\slr}^5/ 16 \pi F^2} .
\label{gravitinodecay}
\eeq
For a given set of weak-scale MSSM parameters leading to a calculable
three-body width for $\slr$, the two-body width Eq.~(\ref{gravitinodecay})
is essentially an independent parameter, depending on $\sqrt{F}$
(or on the gravitino mass in ``no-scale" supergravity models 
\cite{noscalemodels}).
For example, for the sets of parameters and corresponding widths 
in Fig.~2, the three-body decay dominates for $\sqrt{F} \gsim 10^3$ TeV
for $\Delta m - m_{\ell}$
down to a few hundred MeV.
Alternatively, the minimum possible value of $\sqrt{F}$
of order 10 TeV in GMSB 
models corresponds to a maximum width for $\slr \rightarrow \ell\GG$
of order 20 eV (for $m_{\slr}$ of order 100 GeV), so 
$\Delta m$ is expected to be larger
than of order 10 GeV before the three-body decay dominates.
In many of the GMSB models that have actually
been constructed including the supersymmetry-breaking sector
\cite{newGMSBmodels,GMSBmodels}, this limit is not saturated
and $\sqrt{F}$ is orders of magnitude larger than 10 TeV, so the
three-body decay is expected to dominate unless the mass 
difference is correspondingly smaller. Conversely, in ``no-scale" 
models, the two-body decay width might 
even be much larger than the tens of eV range. 

\section*{4. Energy distributions}
\indent 

\begin{figure}[!t]
\centering
\epsfxsize=3.8in
\hspace*{0in}
\epsffile{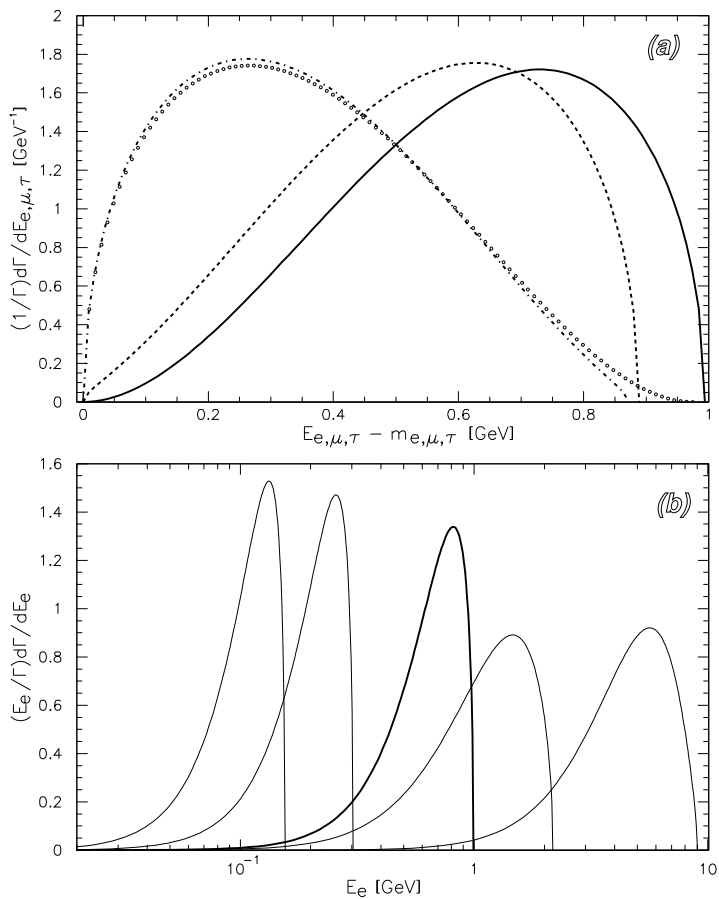}
\caption{Lepton energy distributions in the rest frame of the $\slr^-$  
decaying to $\ell^- \tau^- \stau_1^+$.
(a) Normalized distributions for both the final $\ell$ (solid and 
dashed lines) and $\tau$ (circles and dot-dashed line) for an 
ideal model with 
$N_{11} = 1$; $N_{12} = N_{13} = 0$; $m_{\slr} = 90$ GeV;
$\Delta m = m_{\slr} - m_{\stau_1} - m_\tau = 1$ GeV; $\cos\theta_{\stau} =
0.15$ and $r_{\NI} = 1.1$. Distributions for the other charge channel
are almost identical. The 
solid line and the circles (dashed 
and dot-dashed lines) refer to the case $\ell = e$ ($\ell = \mu$).  
(b) The logarithmic version of the solid thick curve in (a) 
compared to normalized electron-energy distributions in four  
GMSB models chosen from Ref.~\cite{AKMLEP2} (thin lines). 
$\Delta m$ is 0.16, 0.30, 2.2, 9.7 GeV respectively from left to right, 
other details can be found in the text.
}
\label{figdist}
\end{figure}

If the three-body decays of $\ser$ and $\smur$ indeed dominate,
then the $\ell$ and $\tau$ emitted in the decay can be quite
soft if $\Delta m$ is small.  Hence, it is important to address the 
lepton detectability and, in general, the ability to recognize  
a three-body decay pattern in a real experimental environment. 
Using {\tt CompHEP 3.2} \cite{CompHEP} plus an implementation of 
the MSSM lagrangian \cite{Belyaev}, we have examined\footnote{
Note that we have checked in great detail and for a wide
range of parameters that the partial widths for 
three-body $\slr$ decays obtained 
with {\tt CompHEP} are in excellent agreement with
our analytical results given above.}  
the (s)particle energy distributions; those of $e$ or $\mu$ 
and $\tau$ are shown in Fig.~\ref{figdist}(a) and \ref{figdist}(b).
Here, we have plotted the results for 
$\slr^- \rightarrow \ell^- \tau^- \stau^+$, but we have checked 
that the shapes of the normalized distributions
for $\slr^- \rightarrow \ell^- \tau^+ \stau^-$ are essentially identical.
First, we consider 
a model with $\NI = \tilde{B}$, $m_{\slr} = 90$ GeV, $r_{\NI} =  1.1$, 
$\cos\theta_{\stau} = 0.15$, as in the first case of 
Fig.~\ref{figthreebody90}, with $\Delta m = 1$ GeV. 
Fig.~\ref{figdist}(a) shows that the final $e$ or $\mu$ 
(solid thick or dashed line) usually has an energy greater than half 
a GeV in the rest frame of the decaying selectron or smuon. 
Hence, especially when $\slr$ is produced near threshold (as could happen,
e.g.,  
at LEP2) and the boost to the lab frame is small, a successful 
search for the $e$ or $\mu$ in this model requires a detector 
sensitivity at the level of 1 GeV or better (with low associated energy
cuts).
The $\tau$ (circles and dot-dashed line) gets most of the remaining
available energy, so that
$E_\tau - m_\tau$ is usually less 
than 0.5 GeV, while the momentum $|\vec{p}_\tau|$ is usually $\lsim$ 1.5
GeV
in the $\slr$ rest frame. 
It is interesting to note that the final $\stau_1$ can get up to only 
2 GeV in momentum (and usually less), in this case.
In 
the particular model we are considering here, $L$ is of order 
5$\mu$m at LEP2 [from Eq.~(\ref{defL})], and so the kink is impossible to 
detect. However, the decay length could easily be longer in models
with, for example, a larger ratio $m_{\NI}/m_{\slr}$ with fixed external
particle masses. In those cases where the final leptons 
are too soft to be detected, the presence of such a kink in the
charged track might
still signal
a three-body decay pattern.

Most of the above considerations strictly apply to the particular 
model we are considering with $\Delta m = 1$ GeV. Since
the prospects for detection depend crucially on $\Delta m$,
it is important to understand how the distributions scale 
while varying $\Delta m$ (and also other parameters). 
We find that the shapes of the energy distributions in 
Fig.~\ref{figdist}(a) stay basically the same when 
$\Delta m$ is changed, after performing a suitable 
rescaling of the axes.  In addition, we have checked that they  
are only slightly affected by, e.g., changes in $m_{\NI}$ 
and/or stau
mixing angle (within models with $\NI \approx 
\tilde{B}$).
Only when $m_{\NI}$ gets very close to $m_{\slr}$ and/or 
$|\cos\theta_{\stau}| \gsim 0.3$ can deviations exceed 
a few percent (larger deviations are often in the direction 
of shifting the maximum of the $e$ or $\mu$ distribution towards 
slightly lower values, and vice-versa for the tau distribution). 

More generally, in Fig.~\ref{figdist}(b), we illustrate the scaling 
using particular GMSB models from Ref.~\cite{AKMLEP2} that are 
relevant for the slepton three-body decays. We show the logarithmic 
and normalized electron energy distributions for four models (thin
lines) compared to that of Fig.~\ref{figdist}(a) (thick line). 
These four GMSB models have, respectively from left to right: 
$m_{\slr} = 75.8$, 89.8, 63.7, 69.7 GeV; 
\ $\Delta m = 0.16$, 0.30, 2.2, 9.7 GeV; 
\ $\cos\theta_{\stau} = 0.13$; $0.12$, $-0.21$, $-0.31$;  
\ $m_{\NI} = 97.9$, 95.0, 64.6, 75.1 GeV; 
\ $|N_{11}|^2 = 0.88$, 0.97, 0.50, 0.73; 
so that 
\ $\Gamma(\ser^-\rightarrow e^- \tau^- \stau^+_1) = 
$ $9.11 \times 10^{-6}$, $1.11 \times 10^{-3}$, $6.05$, and 237 eV 
\ and \ $\Gamma(\ser^-\rightarrow e^- \tau^+ \stau^-_1) = 
5.19 \times 10^{-6}$, $9.75 \times 10^{-4}$, 5.47 and 171 eV [using 
Eq.~(\ref{fullwidth}) and the corresponding equation for 
$\Gamma(\slr^- \rightarrow \ell^-\tau^- \stau_1^+)$]. 
They were picked in such a way as to probe various 
regions of the GMSB parameter space allowed for models within reach 
of LEP2. 
Fig.~\ref{figdist}(b) shows that, in addition to slight deformations 
of the shapes of distributions due to small $r_{\NI} - 1$ and/or large 
$|\cos\theta_{\stau}| \gsim 0.3$, values of $|N_{11}|^2 \lsim 0.7$
can produce further small changes (as evident from the two models 
more on the right with larger $\Delta m$). 
The total deviations are, however, still small enough to allow 
a model-independent generalization of the discussion above 
concerning the detectability of the three-body decay. 
Thus, it is expected that in most models the $e$ or $\mu$ will typically 
get more than half of the available energy, and hence the chance for 
detection increases straightforwardly with increasing $\Delta m$. 
However, the decay length of the $\ser$ 
or $\smur$ will drop in correspondence with the total width increase, 
diminishing the chance of detecting a 
kink in the charged track.  
Alternatively, for smaller $\Delta m$,  
detection of the $e$ or $\mu$ (and also the $\slr$ kink) is more 
difficult, but of course the decay length is longer, increasing the 
chance that a kink can be seen.

\section*{5. Discussion}
\indent 

At LEP2, the process $e^+e^-\rightarrow \stau^+_1\stau^-_1$
is the most kinematically-favored one for supersymmetry discovery
in the stau NLSP scenario. If the
decay $\stau_1 \rightarrow \tau\GG$ takes place outside the detector
(or inside the detector but with a decay length longer than a few cm),
then the stau tracks (or decay kinks) may be directly
identified \cite{DDRT,DELPHIALEPH}.
If $\ser$ and $\smur$ can also be pair-produced,
then the decays $\slr \rightarrow \ell\tau\stau_1$ studied here can
come into play, leading to additional events 
$e^+e^- \rightarrow \ell^+\ell^- \tau^+\tau^+ \stau_1^- \stau_1^-$ or
$\ell^+\ell^- \tau^-\tau^- \stau_1^+ \stau_1^+$ or
$\ell^+\ell^- \tau^+\tau^- \stau_1^+ \stau_1^-$.
Note that when 
$\Gamma (\slr^- \rightarrow \ell^- \tau^- \stau_1^+)$
is larger than 
$\Gamma (\slr^- \rightarrow \ell^- \tau^+ \stau_1^-)$,
the same-sign $\tau^\pm\tau^\pm \stau_1^\mp \stau_1^\mp$ signals
are suppressed compared to the opposite sign
signals $\tau^+\tau^- \stau_1^+ \stau_1^-$.
In Ref.~\cite{AKMLEP2}, it was observed that the $\smur^+\smur^-$
production
cross section in these models is often significantly larger than
that for $\ser^+\ser^-$, because of the interference effects of
a heavier neutralino in the $t$-channel diagrams contributing to
the latter process. Therefore, one may expect more 
$\mu^+\mu^-\tau\tau\stau_1\stau_1$ events than 
$e^+e^-\tau\tau\stau_1\stau_1$ events, although this is not guaranteed.
We have seen that if $\Delta m$ is smaller than order 1 GeV,
then the identification of soft leptons and taus may be challenging.
However, we noted that in just this case the decay length of $\slr$ 
may well be macroscopic, leading to another avenue for discovery.
Also, since $\slr$ decays isotropically in the rest frame,
and pair-produced sleptons generally do not have a considerable
preference for the beam direction, we expect the probability
for the final particles to be lost down the beam pipe to be small.
This is especially true for $\ell = \mu$, where the production does not
receive contributions from $t$-channel neutralino exchange 
(see, e.g., Ref.~\cite{AKMLEP2}).

If $\stau_1$ decays to $\tau \GG$ with a decay length shorter than a few
cm, then $\stau_1$ decay kinks will be difficult to
observe directly at LEP2. Instead, $\stau^+_1\stau^-_1$ 
production leads only to a signal $\tau^+\tau^-\Etot$.
This has a large background from $W^+W^-$ production, but it may be
possible to
defeat the backgrounds with polar angle cuts \cite{AKMLEP2}.
If $\slr$ pair production is accessible and $\slr\rightarrow\ell\GG$
dominates over $\slr \rightarrow \ell\tau\stau_1$, then the model will
behave essentially like a slepton co-NLSP model, even though the mass 
ordering is naively that of a stau NLSP model. We have seen that this 
might occur even for a multi-GeV $\Delta m$.  Then the most likely 
discovery process may be $e^+e^- \rightarrow \smur^+\smur^- \rightarrow
\mu^+\mu^- \Etot$, as discussed in Ref.~\cite{AKMLEP2}.
If the decay $\stau_1 \rightarrow \tau\GG$ is prompt but 
the decays $\slr \rightarrow
\ell\tau\stau_1$ discussed here still manage to dominate over $
\slr \rightarrow \ell\GG$, then one can have events
$e^+e^- \rightarrow \slr^+\slr^- \rightarrow 
\tau^+\tau^+ (\ell^+\ell^-\tau^-\tau^-)\Etot$ or $
\tau^-\tau^- (\ell^+\ell^-\tau^+\tau^+)\Etot
$ or $\tau^+\tau^- (\ell^+\ell^-\tau^+\tau^-)\Etot$,
with the leptons in parentheses being much softer. The first two
should have very small backgrounds, as will the last one if
the soft leptons are seen.

At the Fermilab Tevatron collider, sleptons can be pair-produced
directly or produced in the decays of charginos
and neutralinos.
If the decays $\stau_1 \rightarrow \tau\GG$ and $\slr \rightarrow
\ell\tau\stau_1$ both take place over macroscopic lengths, then
$p\overline{p} \rightarrow \CI\CI$ or $\CI\NII$ can lead to
events with leptons + jets + heavy charged particle tracks 
(possibly with
decay kinks). It is important to realize that
both the production cross-section and the detection efficiency
for such events will likely be greater than for the direct production
processes $p\overline{p} \rightarrow \slr\slr$ and $\stau_1\stau_1$.
If $\stau_1 \rightarrow \tau\GG$ has a macroscopic decay length but
the decays $\slr \rightarrow \ell\tau\stau_1$ studied here
are prompt, then
there will be some
events with extra soft leptons and taus. However, the latter may be difficult
to detect, and furthermore one may expect that
$\CI$ and $\NII$ will decay preferentially
to $\stau_1\nu_\tau$ and $\stau_1\tau$ (or $\tilde \nu_\tau \tau$ and
$\tilde \nu_\tau \nu_\tau$) rather than through $\slr$.
Similar statements apply for the CERN Large Hadron Collider, except
that the most important source of sleptons may well be from cascade
decays of gluinos and squarks; in some circumstances those decays may
be more likely to contain $\slr$ channels.

In this paper we have studied the three-body decays of selectrons and smuons
in the case that the neutralino is heavier. In GMSB models and other models
with a gravitino LSP, these decays may play a key role in collider 
phenomenology. In particular, we found that the corresponding decay lengths
may be macroscopic and the competition with the decays $\tilde \ell_R
\rightarrow \ell \GG$ may be non-trivial.  We also found that the 
electron or muon in the final state of the three body decay 
usually carries more than half of the available energy in the rest 
frame of the decaying slepton.

{\bf Acknowledgments:} 
We are especially indebted to N.~Arkani-Hamed for leading us to
an important error in an earlier preprint version of this paper.
We are grateful to S.~Thomas and J.~Wells for useful discussions,
and we thank M.~Brhlik and G.~Kane for comments on the manuscript.
S.A. thanks J.~Wells as well as L.~DePorcel and the 1997 SLAC Summer 
Institute staff for providing computer support to pursue preliminary 
work for this paper, while attending the conference.
S.P.M. thanks the Aspen Center for Physics for hospitality.
G.D.K. was supported in part by a Rackham predoctoral fellowship.
This work was supported in part by the U.S. Department
of Energy. 

{\bf Note added, v3 (June 2008):} The earlier v2 of this paper had the 
wrong signs for the coefficients in equations (9) and (10). We are 
grateful to David Sanford, Jonathan Feng, Iftah Galon, and Felix Yu for 
reminding us of this issue. Related signs and figures 3 and 4 have been 
corrected accordingly.



\begin{thebibliography}{99}
\singlespaced

\bibitem{oldGMSBmodels}
    M.~Dine, W.~Fischler, M.~Srednicki, \NPB{189}{1981}{575};
    S.~Dimopoulos, S.~Raby, \NPB{192}{1981}{353};
    M.~Dine, W.~Fischler, \PLBold{110}{1982}{227};
    M.~Dine,  M.~Srednicki, \NPB{202}{1982}{238};
    M.~Dine, W.~Fischler, \NPB{204}{1982}{346};
    L.~Alvarez-Gaum\'e, M.~Claudson, M.~B.~Wise, \NPB{207}{1982}{96};
    C.~R.~Nappi, B.~A.~Ovrut, \PLBold{113}{1982}{175};
    S.~Dimopoulos, S.~Raby, \NPB{219}{1983}{479}.

\bibitem{newGMSBmodels}
    M.~Dine, A.~E.~Nelson, \PRD{48}{1993}{1277};
    M.~Dine, A.~E.~Nelson, Y.~Shirman, \PRD{51}{1995}{1362};
    M.~Dine, A.~E.~Nelson, Y.~Nir, Y.~Shirman, \PRD{53}{1996}{2658}.

\bibitem{Fayet}
    P.~Fayet, \PLBold{70}{1977}{461}; \PLBold{86}{1979}{272};
    \PLB{175}{1986}{471} and in ``Unification of the fundamental 
    particle interactions", eds.~S.~Ferrara, J.~Ellis,   
    P.~van Nieuwenhuizen (Plenum, New York, 1980) p.~587.

\bibitem{oldphotonsignals}
    N.~Cabibbo, G.~R.~Farrar,  L.~Maiani, \PLBold{105}{1981}{155};
    M.~K.~Gaillard, L.~Hall, I.~Hinchliffe, \PLBold{116}{1982}{279};
    J.~Ellis, J.~S.~Hagelin, \PLBold{122}{1983}{303};
    D.~A.~Dicus, S.~Nandi, J.~Woodside, \PLB{258}{1991}{231}.

\bibitem{SWY}
    D.~R.~Stump, M.~Wiest, C.~P.~Yuan, \PRD{54}{1996}{1936}. 

\bibitem{DDRT}    
    S.~Dimopoulos, M.~Dine, S.~Raby, S.~Thomas, 
    \PRL{76}{1996}{3494}; 
    S.~Dimopoulos, S.~Thomas, J.~D.~Wells, \PRD{54}{1996}{3283};
    \NPB{488}{1997}{39}.

\bibitem{AKKMM} 
    S.~Ambrosanio et al., \PRL{76}{1996}{3498}; 
    \PRD{54}{1996}{5395}. 

\bibitem{KKW} 
    K.~S.~Babu, C.~Kolda, F.~Wilczek, \PRL{77}{1996}{3070};
    J.~A.~Bagger, K.~Matchev, D.~M.~Pierce, R.~Zhang, \PRD{55}{1997}{3188}; 
    H.~Baer, M.~Brhlik, C.-h.~Chen, X.~Tata, \PRD{55}{1997}{4463};
    K.~Maki, S.~Orito, hep-ph/9706382;
    A.~Datta et al., hep-ph/9707239;
    Y.~Nomura, K.~Tobe, hep-ph/9708377;
    A.~Ghosal, A.~Kundu, B.~Mukhopadhyaya, hep-ph/9709431.

\bibitem{AKMLEP2} 
    S.~Ambrosanio, G.~D.~Kribs, S.~P.~Martin, \PRD{56}{1997}{1761}. 

\bibitem{DDN} 
    D.~A.~Dicus, B.~Dutta, S.~Nandi, \PRL{78}{1997}{3055};
    hep-ph/9704225; 
    B.~Dutta, S.~Nandi, hep-ph/9709511.

\bibitem{conventions}
    H.~E.~Haber, G.~L.~Kane, \PREP{117}{1985}{75};
    J.~F.~Gunion, H.~E.~Haber, \NPB{272}{1986}{1};
    Erratum, ibid. {\bf B402} (1993) 567.

\bibitem{GMSBmodels} 
    See for example: 
    K.~Intriligator, S.~Thomas, \NPB{473}{1996}{121}; 
    K.-I.~Izawa, T.~Yanagida, \PTP{95}{1996}{829};
    T.~Hotta, K.-I.~Izawa, T.~Yanagida, \PRD{55}{1997}{415};
    E.~Poppitz, S.~Trivedi, \PRD{55}{1997}{5508}; \PLB{401}{1997}{38}; 
    N.~Arkani-Hamed, J.~March-Russell, H.~Murayama, \mbox{hep-ph/9701286};
    S.~Raby, \PRD{56}{1997}{2852}; 
    N.~Haba, N.~Maru, T.~Matsuoka, \PRD{56}{1997}{4207}; 
    C.~Csaki, L.~Randall, W.~Skiba, hep-ph/9707386; 
    S.~Dimopoulos, G.~Dvali, G.~Giudice, R.~Rattazzi, hep-ph/9705307;
    S.~Dimopoulos, G.~Dvali, R.~Rattazzi, hep-ph/9707537; 
    M.~Luty, hep-ph/9706554; 
    M.~Luty, J.~Terning, hep-ph/9709306.
 
\bibitem{noscalemodels} 
    For reviews, see: 
    J.~Kim et al., hep-ph/9707331; 
    J.~L.~Lopez, D.~V.~Nanopoulos, hep-ph/9701264; 
    A.~B.~Lahanas, D.~V.~Nanopoulos, \PREP{145}{1987}{1}.

\bibitem{DELPHIALEPH} 
    The DELPHI collaboration, \PLB{396}{1997}{315}; 
    The ALEPH collaboration, \PLB{405}{1997}{379}; 
    The OPAL collaboration, OPAL PN-306 (1997);
    LEP SUSY Working Group,
    ``Preliminary results from the combination of LEP experiments"
    ({\tt http://www.cern.ch/lepsusy/}).

\bibitem{CompHEP}
    E.~E.~Boos, M.~N.~Dubinin, V.~A.~Ilin, A.~E.~Pukhov, V.~I.~Savrin,
    ``{\tt CompHEP}: Specialized package for automatic calculation of
    elementary particle decays and collisions'', hep-ph/9503280, and
    references therein; P.~A.~Baikov et al., ``Physical results 
    by means of {\tt CompHEP}'', Proc. of X Workshop on High Energy 
    Physics and Quantum Field Theory (QFTHEP-95), eds. B.~Levtchenko, 
    V.~Savrin (Moscow, 1996) p.~101 (hep-ph/9701412).

\bibitem{Belyaev} 
    One of us (S.A.) was provided with a basic code for a supersymmetric
    lagrangian to be implemented in {\tt CompHEP} by A.~Belyaev, 
    to whom we are very grateful. The purpose is to pursue a joint,  
    extended checking program on it and, as part of such a  
    program, we have then checked, corrected and improved the 
    section of the code relevant to this paper, and then used it 
    as indicated. 

\end{thebibliography}
\end{document}